\documentclass[prl,twocolumn]{revtex4}
\usepackage{graphicx}
\usepackage{epsfig}
\usepackage{amsfonts}
\usepackage{amsmath}

\def\be{\begin{eqnarray}}
\def\ee{\end{eqnarray}}

\newcommand{\nn}{\nonumber\\}

\newcommand{\Eq}[1]{Eq.~(\ref{#1})}
\newcommand*{\sd}{^{\dagger}}
\newcommand{\up}{\uparrow}
\newcommand{\down}{\downarrow}

\begin{document}

\title{The Electronic States of Two Oppositely doped Mott Insulators Bilayers}
\author{Tiago C. Ribeiro$^{1,2}$, Alexander Seidel$^{1,2}$, Jung Hoon Han$^{3,4}$
and Dung-Hai Lee$^{1,2}$} \affiliation{${1~}$Department of
Physics,University of California at Berkeley,CA 94720-7300, USA}
\affiliation{${2}$ Materials Sciences Division, Lawrence Berkeley
National Laboratory, Berkeley, CA 94720, USA.} \affiliation{$3$
Department of Physics, Sungkyunkwan University, Suwon 440-746,
Korea}\affiliation{${4}$CSCMR, Seoul National University, Seoul
151-747, Korea}

\date{\today}

\begin{abstract}
We study the effect of Coulomb interaction between two oppositely doped
low-dimensional $tJ$ model systems.
We exactly show that, in the one-dimensional case, an arbitrarily weak 
interaction leads to the formation of charge neutral electron-hole pairs.
We then use two different mean-field theories to address the
two-dimensional case, where inter-layer excitons also form and condense.
We propose that this results in new features which have no analog in single 
layers, such as the emergence of an insulating spin liquid phase.
Our simple bilayer model might have relevance to the physics
of doped Mott insulator interfaces and
of the new four layer Ba$_{2}$Ca$_{3}$Cu$_{4}$O$_{8}$ compound.
\end{abstract}


\maketitle
In recent years, the advance in material science made it possible
to grow atomically smooth epitaxial interfaces between different
iso-structural transition metal oxides \cite{IM9911,OM0278}. In
particular, it is possible to grow atomically ordered (doped) Mott
insulator interfaces \cite{TK8916,BL0373}. Knowing the wealth of
new physics the interface between (doped) band insulators have
produced, and the richer physics of bulk doped Mott insulators as
compared to doped band insulators, the interface science of Mott 
insulators is anticipated to be a promising direction
in years to come. 

Prompted by the above motivation, in this letter we take the
initiative to explore the
interfacial electronic states of two low-dimensional doped Mott insulators. 
We concentrate on the specific case where each Mott insulator is
doped with opposite signed charges, and address the possible 
states caused by the Coulomb attraction between the resulting n and p 
type carriers.
In particular, we find that the formation of electron-hole pairs, 
also known as excitons, across the interface gives rise to new physics 
which is absent in the isolated systems.

More specifically, below we study the interface between either two
one-dimensional (1D) $tJ$ model chains or two two-dimensional (2D)
$tJ$ model layers, where the upper chain (or layer) is doped with 
a density $x$ of electrons and the lower chain (or layer) has the 
same density of holes.
We first consider the 1D case, where we explicitly show that an arbitrarily
small inter-chain Coulomb interaction leads to exciton formation.
We then consider the experimentally relevant 2D system which,
aside from being an explicit model for engineered (doped) Mott 
insulator interfaces \cite{TK8916,BL0373}, may also be pertinent
to address the recently studied stoichiometric four-layer 
Ba$_{2}$Ca$_{3}$Cu$_{4}$O$_{8}$ compound where the simultaneous 
presence of n and p type Fermi surfaces indicates the existence of 
a ``self-doping'' effect \cite{YCHEN}.
Unfortunately, the exact argument used in one dimension does not apply 
in higher dimensions and, in this case, we rather resort two different 
mean-field schemes, namely, a recently developed fully fermionic 
mean-field theory \cite{RW0501,RW0674} and the traditional slave-boson 
mean-field theory \cite{KL8842,WL9603}.
Although these theories sometimes miss the ultimate low energy physics, 
they have successfully predicted the qualitative nature of the cuprate 
phase diagram \cite{WL9603,RW0501}. 
Since our current purpose is to explore the possible
interface electronic phases, we feel that it is meaningful to work
out the prediction of the aforementioned approaches, which we
find yield consistent results.

\textbf{The model.} The Hamiltonian describing our interface
is $H_{interface} = \sum_{a=u,d}H_a + H_{int}$ where
\begin{align}
H_a &= \!\!\!\!\!\! \sum_{\langle ij \rangle \in NN} \!\!\!\!\!
J (\bm{S}_{a,i}.\bm{S}_{a,j} -\frac{n_{a,i} n_{a,j}}{4}) -
t (\tilde{c}_{a,i}\sd \tilde{c}_{a,j} + h.c.)
\label{eq:Ha}
\end{align}
is the Hamiltonian for the upper ($a=u$) and lower ($a=l$) systems
(chains and layers in the 1D and 2D case, respectively).
$J$ and $t$ are the nearest neighbor (NN) antiferromagnetic 
exchange and electron hopping parameters. $\tilde{c}_{a,i}\sd =
\mathcal{P} [c_{a,i,\up}\sd, c_{a,i,\down}\sd] \mathcal{P}$, where
$c_{a,i,\sigma}\sd$ creates electrons in chain (layer) $a$. 
$\mathcal{P}$ projects out doubly occupied (vacant) sites when $a=l$ 
($a=u$). 
$n_{a,i} = \tilde{c}_{a,i}\sd \tilde{c}_{a,i}$ and $\bm{S}_{a,i} =
\tfrac{1}{2} \tilde{c}_{a,i}\sd \bm{\sigma} \tilde{c}_{a,i}$
($\bm{\sigma}$ are the Pauli matrices) are the 
electron number and spin operators.
The across the interface interaction is described by 
\be 
H_{int} = V \sum_i (1-n_{u,i})(1-n_{l,i}) + \frac{\Delta\mu}{2}\sum_i
(n_{u,i}-n_{l,i}) 
\label{eq:Hint} 
\ee 
where $V>0$ is the Coulomb repulsion parameter and $\Delta\mu$ is the 
energy cost to move an electron from the lower to the upper chain (layer).

\textbf{The exactly solvable 1D interface.} 
We first look at the interface model for two 1D chains. 
The model then has the remarkable property
that in the limit $J\rightarrow 0$
the wavefunction of the ground states factorizes
into a spin wavefunction and a charge wavefunction.
This interesting feature was first observed
in Ref. \cite{Ogata1} for the integrable Hubbard chain
at large $U$.
However, it is also common to various non-integrable
$tJ$ type models, leading to a variety of
exact statements \cite{ogata2,seidel}.
We proceed by performing a particle-hole transformation on the upper
chain, which makes doped electrons 
formally identical to holes,
while changing the sign of the interchain interaction, $V\rightarrow -V$.
We then introduce the wavefunction
$\psi(x_1,\sigma_1\dotsc x_N , \sigma_N;\, y_1, \chi_1 \dotsc y_N, \chi_N)$,
where $x_i$, $\sigma_i$ are the positions and spins
of the particles in the upper chain, and $y_i$, $\chi_i$ those of the
lower chain. Both sets of coordinates $x_i$ and $y_i$ are assumed
to be given in ascending order. The action of the Hamiltonian on
 the wavefunction $\psi$ now takes the following form
\begin{eqnarray}\label{eq:H}
H\psi=&-t\sum_{l=1}^N(\psi(\dotsc x_l\pm 1\,\sigma_l\dotsc)+\psi(\dotsc y_l\pm 1\,\chi_l\dotsc))\nn
&-V\sum_{l=1}^N\sum_{m=1}^N \delta_{x_l,y_{m}}\;\psi(\dots)+H_J\;\psi(\dots)
\end{eqnarray}
where $H_J$ represents the spin exchange term proportional to $J$.
In the limit $J\rightarrow 0$ one can easily verify that the eigenstates
of the system described by \Eq{eq:H} are of the following factorized
form:
\begin{equation}\label{eq:factor}
\psi(\dotsc)= f(x_1\dotsc x_N;y_1\dotsc y_N)\,
g(\sigma_1\dotsc \sigma_N; \chi_1\dotsc \chi_N).
\end{equation}
This follows from the fact that at $J=0$, the Hamiltonian \Eq{eq:H}
does not act on the spin degrees of freedom. Hence, $f$ must be
an eigenfunction of \Eq{eq:H} with $J=0$, whereas $g$ can be an
{\em arbitrary} function of spin variables. Precisely at $J=0$
the ground state is thus hugely degenerate in the spin sector.
This degeneracy can be lifted in first order perturbation theory
via an effective spin Hamiltonian acting on a spin ladder
of $2N$ spins, where $N$ is the number of singly occupied sites in
both the upper and the lower chain, and the holes are effectively
``squeezed'' out of the system \cite{ogata2, seidel}. At first order
in $J$, the spin wavefunction $g$ is thus uniquely determined. Here
we are, however, more concerned with the charge correlations of
the system. In this regard, we observe that the determination of
the charge wavefunction $f$ from the Hamiltonian
\Eq{eq:H} is equivalent to finding the ground state of the
attractive Hubbard model. Indeed, the effective charge
Hamiltonian acting on $f$ can be rewritten in second quantized form as
\begin{equation}\label{eq:Hub}
  H_c=-t\sum_{i,a}(a^\dagger_{i+1,a}a_{i,a}+ h.c.) 
-V\sum_i a^\dagger_{i,u}a_{i,u} a^\dagger_{i,d}a_{i,d}
\end{equation}
where the fermion operator $a_{i,a}$ carries a ``pseudospin''
index $a$ labeling the layer. It is well known that \Eq{eq:Hub} is in the spin gapped
Luther-Emery liquid phase of 1d systems for any $V>0$, which is
the 1D analog of a paired superfluid \cite{seidel}.
The spin gap of the Hubbard model hence implies a charge gap due
to the formation of electron-hole pairs in the double-chain system.
The important lesson to be learned is that, in
one dimension, the phase diagram of our interface model features
exciton formation at {\em arbitrarily weak} inter-chain interaction.

\textbf{The fully fermionic mean-field theory.} We now discuss the 
2D bilayer Hamiltonian $H_{interface}$ at the mean-field 
level. We start with the fully fermionic mean-field theory 
introduced in Refs. \onlinecite{RW0501,RW0674}.
This theory introduces fermionic operators $d_{a,i}\sd = [d_{a,i,\up}\sd
d_{a,i,\down}\sd]$ that create charge $(+e)$ and spin-1/2 holes in
the lower Hubbard band of the lower layer ($a=l$) and charge $(-e)$
and spin-1/2 electrons in the upper Hubbard band of the upper
layer ($a=u$).
The Mott insulating background on top of which these carriers move
is described by a singly occupied lattice of neutral spin-1/2
fermions whose creation operators are $f_{a,i}\sd = [f_{a,i,\up}\sd
f_{a,i,\down}\sd]$.
In terms of the $d$ and $f$ fermions the projected electron operators 
can be expressed as \cite{RW0674}
\begin{align}
\tilde{c}_{l,i}\sd &= s_{\sigma} \frac{1}{\sqrt{2}}
\left[\left( \frac{1}{2} + s_{\sigma} \widetilde{S}_{l,i}^z \right)
\tilde{d}_{l,i,-\sigma} - \widetilde{S}_{l,i}^{s_{\sigma}} 
\tilde{d}_{l,i,\sigma} \right] \notag \\
\tilde{c}_{u,i}\sd &= \frac{1}{\sqrt{2}}
\left[\left( \frac{1}{2} - s_{\sigma} \widetilde{S}_{u,i}^z \right)
\tilde{d}_{u,i,\sigma}\sd - \widetilde{S}_{u,i}^{s_{\sigma}} 
\tilde{d}_{u,i,-\sigma}\sd \right]
\label{eq:electron_enl}
\end{align}
where $s_{\sigma} = (+1),(-1)$ for $\sigma = \up,\down$,
$\widetilde{\bm{S}}_{a,i}= \tfrac{1}{2} f_{a,i}\sd \bm{\sigma}
f_{a,i}$ and 
$\tilde{d}_{a,i,\sigma} = d_{a,i,\sigma} 
(1 - d_{a,i,-\sigma}\sd d_{a,i,-\sigma})$.
These relations allow us to recast $H_{interface}$ in terms of the
$d,d\sd$ and $f,f\sd$ operators, after what we perform a mean-field
factorization to obtain a quadratic Hamiltonian.

First, we summarize the (single layer) mean-field results of Ref.
\cite{RW0501,RW0674}.
At low enough doping the $tJ$ model develops antiferromagnetic 
order \cite{M9101}, which yields non-zero staggered magnetization 
order parameters
\begin{align}
m &= \frac{1}{2} s_a (-)^{i_x+i_y}\langle \psi_{a,i}\sd \psi_{a,i} - 1
\rangle \notag \\
n &= -s_a (-)^{i_x+i_y} \langle\sum_{\nu=2,3} t_{\nu} \sum_{\hat{u}\in \nu \,
NN} \eta_{a,i}\sd \eta_{a,i+\hat{u}}+h.c.\rangle
\label{eq:mfp_af}
\end{align}
Here, $\psi_{a,i}\sd =
[\psi_{a,i,1}\sd \psi_{a,i,2}\sd] = s_a^{i_x-i_y}[f_{a,i,\up}\sd
f_{a,i,\down}]$ and $\eta_{a,i}\sd = [\eta_{a,i,1}\sd
\eta_{a,i,2}\sd ] = s_a^{i_x-i_y}[d_{a,i,\up}\sd d_{a,i,\down}]$;
$  s_a=(+1),(-1)$ for $a=l,u$; $t_2=J$ and $t_3=J/2$;
$\hat{u} = \pm\hat{x}\pm\hat{y}$ for $\nu=2$ and $\hat{u} =
\pm2\hat{x},\pm2\hat{y}$ for $\nu=3$ \cite{KSHIFT}. In addition to
the antiferromagnetic order parameter, singlet bond order
parameters which capture the spin liquid correlations are also
introduced  
\be \chi = \langle\psi_{a,i}\sd \sigma_z
\psi_{a,j} \rangle \ ; \ \ \Delta =
(-)^{i_y-j_y}\langle\psi_{a,i}\sd \sigma_x \psi_{a,j} \rangle
\label{eq:mfp_rvb} 
\ee 
with $\langle ij\rangle\in NN$. Finally,
the Bose amplitudes which describe the quantum coherence between
the doped carriers and the spin background are given by \be b_0 =
\langle f_{a,i}\sd d_{a,i}\rangle \ ; \ \ b_1 =
\langle\frac{3}{16}\sum_{\nu = 1,2,3} t_{\nu}\sum_{\hat{u}\in \nu
\, NN} f_{a,i}\sd d_{a,i+\hat{u}}\rangle \label{eq:mfp_int} \ee
where $t_1=t$ and $\hat{u} = \pm\hat{x},\pm\hat{y}$ for $\nu=1$.

In terms of these order parameters, the in-plane mean-field
Hamiltonian is given by \cite{RW0501,RW0674}
\begin{align}
H_a^{MF} &=  \sum_{a,\bm{k}} \left[ \begin{array}{cc}
\psi_{a,\bm{k}}\sd~\eta_{a,\bm{k}}\sd \end{array} \right] \left[
\begin{array}{cc} \alpha_{\bm{k}}^z \sigma_z + \alpha_{\bm{k}}^x
\sigma_x &
\beta_{\bm{k}}\sigma_z \\
\beta_{\bm{k}} \sigma_z & \gamma_{\bm{k}}^z \sigma_z \end{array}
\right] \left[ \begin{array}{c}
\psi_{a,\bm{k}} \\ \eta_{a,\bm{k}} \end{array} \right] \notag \\
&+ \sum_{a,\bm{k}} \left( \nu_a^{\psi} \psi_{a,\bm{k}+\bm{Q}}\sd
\psi_{a,\bm{k}} + \nu_{a,\bm{k}}^{\eta} \eta_{a,\bm{k}+\bm{Q}}\sd
\eta_{a,\bm{k}} \right) 
\label{eq:HMF}
\end{align}
In the above equation $\alpha_{\bm{k}}^z =
-(\tfrac{3\tilde{J}}{4}\chi - t_1 x ) (\cos k_x+\cos k_y)+a_0$
where $a_0$ is the Langrange multiplier that ensures $\langle
f_{a,i}\sd f_{a,i} \rangle = 1$, and $\tilde{J} = (1-x)^2 J$. In
addition, $\alpha_{\bm{k}}^x = - \tfrac{3\tilde{J}}{4} \Delta
(\cos k_x-\cos k_y)$, $\beta_{\bm{k}} = -\tfrac{3b_0}{8} [t_1(\cos
k_x+\cos k_y) + 2 t_2\cos k_x \cos k_y + t_3 (\cos 2k_x+\cos
2k_y)] + b_1$, $\gamma_{\bm{k}}^z = t_2 \cos k_x \cos k_y +
\tfrac{t_3}{2} (\cos 2k_x+\cos 2k_y) + \tfrac{\Delta\mu}{2}$,
$\nu_a^{\psi} = 2 s_a (n - 0.34\tilde{J}m)$,
$\nu_{a,\bm{k}}^{\eta} = 2 s_a m (\tfrac{\Delta\mu}{2} -
\gamma_{\bm{k}}^z)$. ,

Unlike other slave-particle approaches 
\cite{KL8842,WL9603,RV8893}, all particles in the above mean-field
theory are fermions. The central advantage of this mean-field
approach is the ease to describe fermionic quasiparticles, which are
created by the $d\sd$ operator.
The ``holon'' in the slave-boson approach, which is a spin singlet 
and charge $+e$ non-quasiparticle object, is the composite bosonic 
particle made up of a $d$ quasiparticle and a $f$ neutral fermion, and its 
condensation is described by the the order parameters $b_{0,1}$ \cite{RW0674}.
The fully fermionic approach, which yields results consistent 
with numerical calculations \cite{RW0674},
successfully reproduces various non-trivial aspects
of ARPES and STM experiments in the cuprate superconductors, including
the Fermi arcs, the dispersion kink, the minimum gap 
locus, the flat dispersion around $(\pi,0)$, the peak-dip-hump and 
the absence of coherence peaks in deeply underdoped $dI/dV$ spectra 
\cite{RW0501,RW0603}.
Encouraged by the above successes we extend \Eq{eq:HMF} to
describe two oppositely doped Mott insulator layers. 
Following our exact 1D result, we introduce the order parameters
\begin{align}
e_s &= \langle \eta_{l,i}\sd \sigma_- \eta_{u,i} + \eta_{u,i}\sd \sigma_-
\eta_{l,i} \rangle \notag \\
e_t &= (-)^{i_x+i_y} \langle \eta_{l,i}\sd \sigma_- \eta_{u,i} -
\eta_{u,i}\sd \sigma_- \eta_{l,i} \rangle, \label{eq:mfpexc}
\end{align}
to describe the formation of spin singlet and spin triplet
excitons. In \Eq{eq:mfpexc} $\sigma_- = (\sigma_x-i\sigma_y)/2$.
The resulting inter-layer mean-field Hamiltonian reads
\begin{align}
H_{int}^{MF} &= - \frac{V}{2} \sum_{\bm{k}} \left( e_s \eta_{u,\bm{k}}\sd
\sigma_x \eta_{l,\bm{k}} + i e_t \eta_{u,\bm{k}+\bm{Q}}\sd
\sigma_y \eta_{l,\bm{k}} + h.c. \right)
\label{eq:HMF_int}
\end{align}

\begin{figure}
\begin{center}
\includegraphics[width=0.49\textwidth]{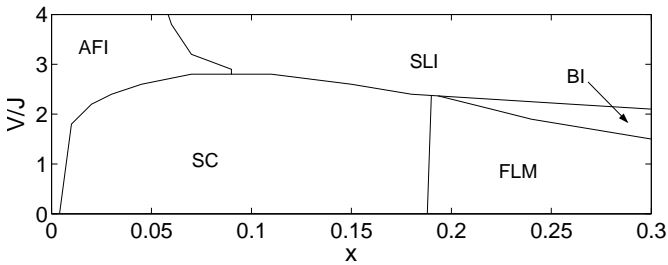}
\caption{\label{fig:pdiagram}
$tJ$ model bilayer mean-field phase diagram for
$t=3J$.
It includes the antiferromagnetic insulator (AFI), the spin liquid 
insulator (SLI), the superconductor (SC), the band insulator (BI) and 
the Fermi liquid metal (FLM) described in the main text.
}
\end{center}
\end{figure}

 The zero temperature
mean-field phase diagram is determined upon minimizing the
mean-field energy $\langle H_u^{MF} + H_l^{MF} + H_{int}^{MF} -
E_0 \rangle$, where $E_0 = - [\tfrac{3\tilde{J}N}{2}
(\chi^2+\Delta^2) + 1.36\tilde{J}Nm^2 - 8Nmn - 4Nb_0b_1 +
\tfrac{V}{2}N (e_S^2+e_T^2) + \Delta\mu N]$ and $N$ is the number
of lattice sites. Motivated by the mutual exclusion of
antiferromagnetism and superconductivity \cite{BL0373} in the
cuprates, we omit the possible coexistence of antiferromagnetic 
and superconducting orders.
The resulting phase diagram for the bilayer $t=3J$ model is depicted in
Fig. \ref{fig:pdiagram}.
It comprises five distinct regions 
whose properties we discuss below:

\textit{(i) AF insulator} ($\chi,m,n,e_s,e_t \neq 0$ and
$b_0,b_1 = 0$).
Since $m\neq 0$ spins are antiferromagnetically ordered.
Electrically neutral excitons form ($e_s,e_t \neq 0$) which
cost a finite energy to break, thus opening a gap in the
charge sector.

\textit{(ii) Gapless spin liquid insulator} ($\chi,\Delta,e_s
\neq 0$ and $m,n,e_t,b_0,b_1 = 0$).
In the absence of AF order ($m=0$), the spins are in a gapless $d$-wave
($\Delta \neq 0$) paired spin liquid state reminiscent of the
pseudogap state in the single layer problem \cite{WL9603}.
However, in the bilayer case excitons ($e_s \neq 0$) render the system
a charge insulator.

\textit{(iii) $(D+S)$-wave superconductor}
($\chi,\Delta,b_0,b_1,e_s \neq 0$ and $m,n,e_t=0$). Doped carriers
hybridize with the $d$-wave paired spins ($\Delta,b_0 \neq 0$) and,
at the same time, participate in forming $s$-wave excitons. As a
result, the superconducting gap acquires a $D+S$ symmetry.

\textit{(iv) Band insulator} ($\chi,b_0,b_1,e_s \neq 0$ and
$m,n,e_t,\Delta = 0$). In the absence of pairing ($\Delta=0$)
there is no superconducting order. The non-zero $b_0$ and $e_s$ 
condensates open a gap in both the spin and charge
sectors and the system behaves as a renormalized band insulator.

\textit{(v) Fermi liquid metal} ($\chi,b_0,b_1 \neq 0$ and
$m,n,e_t,e_s,\Delta = 0$). In the absence of $\Delta$ and $e_s$
the quasiparticle gap closes and the system behaves as a
renormalized Fermi liquid.

The phase diagram in Fig. \ref{fig:pdiagram} results from the
competition between the spin exchange energy, the kinetic energy
(of the doped carriers) and the inter-layer Coulomb energy. The
latter two vanish at half-filling, hence, at sufficiently low
doping, antiferromagnetic order prevails. As doping increases,
the itinerancy of charge carriers tends to destroy antiferromagnetism 
and to favor a spin liquid state instead \cite{R0637}. 
In that case, and for sufficiently large $V$, the antiferromagnetic
insulator is replaced by the spin liquid insulator where all doped 
carriers form inter-layer excitons. 
Since charge carriers are more mobile when they are not part of 
an exciton, partial exciton unbinding is favored at low $V$. 
The carriers thus liberated from the excitonic condensate then
bind to the spins and, in the presence of spin pairing,
the system is a superconductor. 
At large doping, kinetic energy dominates over the exchange 
interaction and, as a result, the superconducting gap closes, 
rendering the bilayer metallic.
This metallic state occurs in a system with two electrons per 
bilayer unit cell and, nominally, it is rather expected to 
be an insulator.
Such band theory expectations are amiss due to the effect
of strong correlations \cite{GK0603}.
Since the effect of electron-electron correlations decreases
upon further doping, the system eventually becomes a band 
insulator, as expected in the weak coupling limit.

\textbf{The slave-boson mean-field theory.} To vindicate the above
predictions we also determine the phase diagram using the 
slave-boson mean-field theory \cite{KL8842,WL9603}.
We use the antiferromagnetic $(m)$, in-plane $d$-wave pairing
$(\Delta^f )$, inter-plane exciton pairing $(\Delta^b)$ and single
boson condensation $(\langle b \rangle )$ 
mean-field order parameters and, as before, ignore the coexistence
of antiferromagnetic and superconducting orders. To our
satisfaction, the results are in qualitative agreement with those
obtained using the above fully fermionic mean-field theory. 
Upon increasing the carrier concentration, the 
excitonic antiferromagnetic insulator ($m\neq 0$, $\Delta^b \neq0$,
$\Delta^f \neq 0 $, $\langle b\rangle = 0)$
present at $ 0 < x < x_c $ evolves to the excitonic 
$d$-wave superconductor
($m = 0$, $\Delta^b \neq 0$,  $\Delta^f \neq 0$, $\langle b
\rangle\neq 0$) for $x > x_c$. The critical concentration $x_c$
increases with $V/J$ in a manner consistent with the AFI-SC phase
boundary depicted in Fig. \ref{fig:pdiagram}.
If $V/J$ is larger than a certain value, 
however, we find that the phase that replaces the AF state is a 
spin liquid insulator ($m = 0$, $\Delta^f \neq 0$, $\Delta^b \neq 0$,
$\langle b \rangle = 0$). 
A similar AFI-to-SLI transition is found for large $V/J$ values in
Fig. \ref{fig:pdiagram}.
Detailed findings of the slave-boson calculation are presented
elsewhere \cite{HJ0626}.

\textbf{Discussion.} Above, we use exact arguments and two different
mean-field theories to conclude that Coulomb interaction between 
oppositely doped $tJ$ model systems leads to the formation of excitons
in 1D and 2D, respectively.
Since in the 1D [2D] system the hole momentum distribution is shifted 
from that of electrons by $\pi$ [$(\pi,\pi)$], excitons 
carry a finite momentum equal to $\pi$ [$(\pi,\pi)$] \cite{KSHIFT}.

Even though in the 2D case our model Hamiltonian $H_{interface}$ is, 
at most, a simplified version of material compounds' Hamiltonians, 
it provides the playground to explore the important effect of inter-layer
Coulomb interaction, which is the largest interaction between 
neighboring oppositely doped Mott insulators.
We also believe that the single layer physics is correctly captured 
by restricting the in-plane Coulomb interaction to electrons sharing
the same lattice site.
However, it is well known that terms describing electron hopping 
beyond NN sites can significantly alter the single plane properties
\cite{R0637,TM9496}.
Above, we neglect such terms, which may limit our 
ability to address systems like the Ba$_{2}$Ca$_{3}$Cu$_{4}$O$_{8}$ 
compound where farther neighbor hopping processes are believed to 
be important. 
Although inclusion of longer range hopping is the subject of 
future work, we expect the present conclusions to apply as long as 
$xt/t',xV/t' \gtrsim 1$.

Finally, we note that $H_{interface}$ is interesting in itself, as it 
illustrates how exciting new physics (absent in isolated single layers) 
may emerge in bilayer systems.
In the present case, such physics is summarized as follows:

\textit{(i)} In the presence of large inter-layer Coulomb 
interaction all doped carriers form electrically neutral 
electron-hole pairs and the system is an insulator. 
As these pairs move throughout the layers,
they disorder the spin background and stabilize a fractionalized
spin liquid state which supports neutral spin-1/2 excitations.
So far, there is no evidence for such spin liquid states in single 
layers \cite{BW0187}, where the itinerant charge carriers render 
the spin liquid unstable to superconductivity at low temperature. 
Here, we propose that the charge gap in the excitonic insulating 
phase can protect the exotic spin liquid state in the bilayer formed 
by oppositely doped Mott insulators. 

\textit{(ii)} The coexistence of the excitonic condensate and 
superconductivity changes the quasiparticle gap from pure $D$ to 
$D+S$ symmetry. 
The extra gap component deviates the nodes away from the 
$(0,0)-(\pi,\pi)$ and $(0,0)-(-\pi,\pi)$ directions. 
This effect can, in principle, be detected in ARPES experiments.

TCR was supported by the FCT Grant No. SFRH/BPD/21959/2005 (Portugal) and
the DOE Grant No. DE-AC02-05CH11231.
AS and DHL acknowledge support by the DOE Grant DE-AC03-76SF00098.
JHH was supported by the KRF Grant No. KRF-2005-070-C00044 (Korea).

\newcommand*{\PRL}[1]{Phys.\ Rev.\ Lett.\ {\textbf {#1}}}
\newcommand*{\PRB}[1]{Phys.\ Rev.\ B {\textbf {#1}}}
\newcommand*{\RMP}[1]{Rev.\ Mod.\ Phys.\ {\textbf {#1}}}
\newcommand*{\CMAT}[1]{cond-mat/{#1}}
\newcommand*{\SST}[1]{Supercond.\ Sci.\ Technol.\ {\textbf {#1}}}
\newcommand*{\NA}[1]{Nature\ {\textbf {#1}}}


\begin{thebibliography}{99}
\bibitem{KSHIFT} We shift the momenta of the upper layer
operators by $(\pi,\pi)$ to make $H_u$ formally equivalent to $H_l$.
\end{thebibliography}
\end{document}